\documentclass[aps,twocolumn,prb]{revtex4}

\usepackage{graphicx}
\usepackage{dcolumn}
\usepackage{bm}

\begin{document}
\preprint{APS/123-QED}

\hspace{200mm}
\vspace{7mm}
\title{Conventional $s$-wave superconductivity in BiS$_2$-based NdO$_{0.71}$F$_{0.29}$BiS$_{2}$ revealed by thermal transport measurements}

\author{T. Yamashita$^1$}
\author{Y. Tokiwa$^2$}
\author{D. Terazawa$^1$}
\author{M. Nagao$^3$}
\author{S. Watauchi$^3$}
\author{I. Tanaka$^3$}
\author{T. Terashima$^2$}
\author{Y. Matsuda$^1$}

\affiliation{
$^1$Department of Physics, Graduate School of Science, Kyoto University, Kyoto 606-8502, Japan\\
$^2$Research Center for Low Temperature and Materials Science, Kyoto University, Kyoto 606-8501, Japan\\
$^3$Center for Crystal Science and Technology, University of Yamanashi, Kofu 400-8511, Japan
}

\date{\today}

\begin{abstract}
To study the superconducting gap structure of BiS$_2$-based layered compound NdO$_{0.71}$F$_{0.29}$BiS$_{2}$ ($T$$_c$ = 5 K), we measured the thermal conductivity $\kappa$,  which is a sensitive probe of low-energy quasiparticle spectrum.    In the absence of a magnetic field, residual linear term  in the thermal conductivity $\kappa_{0}$/$T$ at $T$ $\rightarrow$ 0 is vanishingly small,  indicating that the residual normal fluid, which is expected for nodal superconductors, is absent.   Moreover,  the applied magnetic field hardly affects thermal conductivity in wide range of the vortex state, indicating the absence of Doppler shifted quasiparticles.  These results provide evidence that NdO$_{0.71}$F$_{0.29}$BiS$_{2}$ is a fully gapped superconductor.  The obtained gap structure, along with the robustness of the superconductivity against the impurity, suggest a conventional $s$-wave superconducting state in  NdO$_{0.71}$F$_{0.29}$BiS$_{2}$.
\end{abstract}
\maketitle

Recently a new family of layered superconductors Bi$_4$O$_4$S$_3$ and $Ln$O$_{1-x}$F$_x$BiS$_2$ ($Ln$ is a lanthanoid)~\cite{Mizuguchi-prb12,Mizuguchi-jpsj12,Xing-prb12,Demura-jpsj13,Yazici-PM13,Jha-JSNM13} has been reported.  Superconductivity emerges from semiconducting parent compound via electron doping by substituting O with F in the blocking layer.  Up to now the highest  transition temperature $T_c$ of 10.6 \,K is reported  in LaO$_{0.5}$F$_{0.5}$BiS$_2$~\cite{Mizuguchi-jpsj12}.  The crystal structure consists of alternate stacking of BiS$_2$ superconducting double layers and $Ln$O insulating blocking layers (inset of Fig.\,1(a)). Fermi surfaces consist of two-dimensional (2D) cylindrical sheets. The band structure calculations suggest the presence of strong Fermi surface nesting at the wave vector $(\pi,\pi)$~\cite{Usui-prb12,Wan-prb13,Yildirim-prb13,Agatsuma-JMMM16}.  In particular, in the underdoped regime ($x<0.5$), the Fermi surface consists of disconnected small electron pockets at the Brillouin Zone boundary  (inset of Fig.\,1(b)).  Because of some common features with Fe-based high temperature superconductors,  BiS$_2$-based superconductors have aroused great interest.  On the other hand,  the conduction band in the 2D layer of BiS$_2$-based compounds is mainly of 6$p_x$ and 6$p_y$ orbits of Bi~\cite{Usui-prb12}.  The electron correlation effects, which play an essential role for the superconductivity in Fe-based superconductors, appear not to be important due to the widely spread 6p orbitals.   Thus a major outstanding question is whether the Cooper pairing is mediated by  unconventional (non-phononic)  interactions, such as antiferromagnetic fluctuations.  To elucidate this issue, the identification of  superconducting gap structure  is of primary importance, because it is  intimately related to the pairing interaction.  

Several superconducting gap structures, including conventional $s$-, sign reversing $s$-, spin triplet $p$-, and $d$-wave symmetries have been proposed for BiS$_2$-based superconductors theoretically~\cite{Usui-prb12,Wan-prb13,Yildirim-prb13,Liang-FP14,Agatsuma-JMMM16,Yang-prb13,Martins-prb13}.  A possible unconventional superconductivity has been reported by an extremely large ratio of  $2\Delta/k_BT_c\sim$17 ($\Delta$ is the superconducting gap), which is nearly five times larger than the BCS value~\cite{Liu-EPL14}. Fully gapped superconductivity has been reported by $\mu$SR measurements of polycrystalline Bi$_4$O$_4$S$_3$ and LaO$_{0.5}$F$_{0.5}$BiS$_2$~\cite{Lamura-prb13}. To look into the superconducting state, the measurements on single crystals are more desirable. Raman scattering experiments on single crystals of NdO$_{1-x}$F$_x$BiS$_2$ with $x\sim$0.5 have suggested a possible phonon-mediated superconductivity~\cite{Wu-prb14}. Moreover, recent measurements of London penetration depth $\lambda(T)$ on single crystals for $x$=0.3 and 0.5 have also reported the fully gapped superconductivity~\cite{Jiao-JPhys15}. However, the large Curie-Weiss contribution at low temperature arising from localized Nd$^{3+}$ spins  prevents the accurate determination of superfluid density by the observed $\lambda$.

Here, to provide conclusive information on the superconducting gap structure of BiS$_2$-based superconductors,  we performed thermal conductivity  measurements on a high quality single crystal of NdO$_{0.71}$F$_{0.29}$BiS$_{2}$ down to 100\,mK. It is well established that low-temperature thermal conductivity is a powerful probe for superconducting gap structure, detecting low-energy quasiparticle excitations sensitively~\cite{Matsuda-JPhys06}. Advantage of thermal conductivity is that it is insensitive to Nd magnetic moments. By thermal conductivity measurements, we provide evidence that NdO$_{0.71}$F$_{0.29}$BiS$_{2}$ is a fully gapped superconductor. Based on these results, together with the impurity effect, we discuss the mechanism of superconductivity. 

NdO$_{0.71}$F$_{0.29}$BiS$_{2}$ single crystals  were grown by the high-temperature flux method with CsCl/KCl as a flux\cite{Nagao-jpsj13}.  The onset of superconductivity is $T$$_{\rm{c}}$$^{\rm{onset}}$ $\sim$ 5.2 K, which is in agreement with the previous report\cite{Nagao-jpsj13}. The superconducting volume fraction determined  by the magnetic susceptibility measurements is close to 100 \%.   The X-ray diffraction shows no impurity phase. Thermal conductivity $\kappa$ was measured along the tetragonal $a$ axis (heat current {\boldmath $q$}$\parallel a$) on a sample with a rectangular shape ($\sim$2.6$\times$1.16$\times$0.015 mm$^3$) by the standard steady state method in a $^3$He-$^4$He dilution refrigerator.   

\begin{figure}[t]
	\vspace{12mm}
	\includegraphics[width=0.8\linewidth,keepaspectratio]{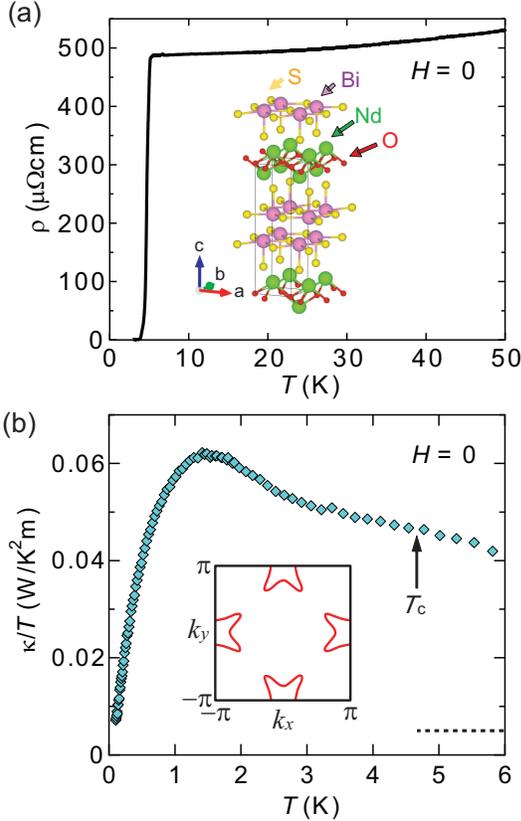}
	\caption{(Color online) Temperature dependence of (a) electrical resistivity $\rho$ and (b) thermal conductivity divided by temperature $\kappa/T$ of NdO$_{0.71}$F$_{0.29}$BiS$_{2}$ at zero field. Dashed line indicates the electronic contribution estimated from the Wiedemann-Franz law. Inset of (a) displays the crystal structure of NdO$_{1-x}$F$_x$BiS$_2$. Inset of (b) shows Fermi surface of NdO$_{1-x}$F$_x$BiS$_2$ for $x\sim 0.3$~\cite{Usui-prb12,Agatsuma-JMMM16}.}
\end{figure}

Figure~1(a) shows the temperature dependence of the in-plane electrical resistivity $\rho (T)$ of  NdO$_{0.71}$F$_{0.29}$BiS$_{2}$.  Below 50\,K, $\rho (T)$ exhibits Fermi liquid behavior with $T^2$-dependence down to $T_c$. Figure 1(b) shows the $T$ dependence of $\kappa/T$ in zero field.  At $T_c$,  $\kappa/T$ shows no discernible anomaly. The electronic contribution estimated by assuming the Wiedemann-Franz law, $\kappa/T=L_0/\rho$, where $L_0=\frac{\pi^2}{3}\left(\frac{k_B}{e}\right)^2$ is the Lorenz number,  is about 11 \% of the total thermal conductivity at $T_c$.   In the superconducting state,  $\kappa/T$ decreases with decreasing $T$ after showing a broad maximum at around 1.5\,K. We will discuss this behavior later.

We first discuss the low temperature behavior of $\kappa/T$ in zero field. Thermal conductivity can be written as a sum of the quasiparticle and phonon contributions, $\kappa=\kappa_{qp}+\kappa_{ph}$.   The phonon conductivity in boundary-limited scattering regime at low temperature is expressed as
\begin{equation}
\kappa_{ph} = \frac{1}{3} \beta \langle v_{s} \rangle \ell_{ph} T^{3},
\end{equation}
where $\beta$ is the phonon specific heat coefficient, $\langle v_{s} \rangle$ is the mean acoustic phonon velocity, and $\ell_{ph}$ is the phonon mean free path.  For diffuse scattering limit, $\ell_{ph}$ becomes $T$-independent, resulting in $\kappa_{ph}\propto T^3$.  On the other hand, in case of specular reflection, $\ell_{ph}$ follows $T^{-1}$-dependence, leading to $\kappa_{ph}\propto T^2$. In real systems, $\kappa_{ph}\propto T^{\alpha}$ with $\alpha$ of intermediate value between 2 and 3.  In fact, $\alpha=$2.74, 2.4 and 2.77 has been reported in V$_3$Si~\cite{Sutherland-prb03}, YBa$_2$Cu$_3$O$_{6.99}$~\cite{Sutherland-prb03} and Al$_2$O$_{3}$~\cite{Pohl-prb82}, respectively. 

\begin{figure}[t]
	\vspace{13mm}
	\includegraphics[width=\linewidth,keepaspectratio]{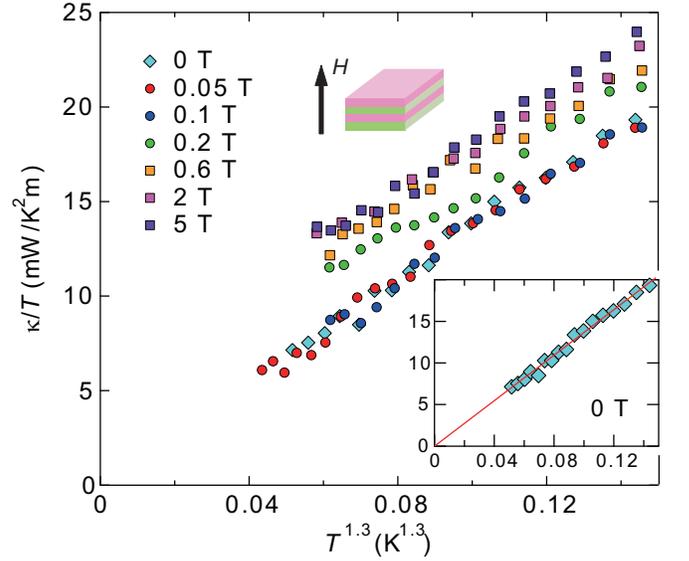}
	\caption{(Color online) Thermal conductivity divided by temperature $\kappa/T$ of NdO$_{0.71}$F$_{0.29}$BiS$_{2}$ plotted against $T^{1.3}$ in several magnetic fields applied parallel to the c-axis. Inset: $\kappa/T$ vs $T^{1.3}$ in zero field. Solid line represents linear fit to the data below 0.3\,K.}
\end{figure}

We found that  $\kappa(T)$ is well fitted as $\kappa\propto T^{2.2-2.5}$ in the widest temperature range of $T<$ 300\,mK.  In the inset of Fig.\,2, $\kappa/T$  in zero field is plotted as a function of $T^{1.3}$.  As shown with the red line,  $\kappa/T$ is extrapolated to zero at $T\rightarrow 0$, i.e.  the absence of a residual term.  The vanishingly small residual term is obtained in the range of $\alpha$ between 2.2 and 2.5.   We note that  for $\alpha=3$,  there is a finite residual term but well fitted range is limited to $T<150$\,mK and for $\alpha=2$, the residual term becomes negative. It is well known that finite residual term indicates existence of a residual normal fluid, which is expected for an unconventional superconductor with line nodes in the energy gap.  This residual normal fluid is a consequence of impurity scattering, even for low concentrations of nonmagnetic impurities.  	 Therefore the absence of residual term  suggests that there is no line nodes in NdO$_{0.71}$F$_{0.29}$BiS$_{2}$.  Since the presence of point node is unlikely in the 2D Fermi surface, the zero field thermal conductivity suggest a fully gapped superconductivity.  
	

\begin{figure}[t]
      	\vspace{16mm}
	\includegraphics[width=\linewidth]{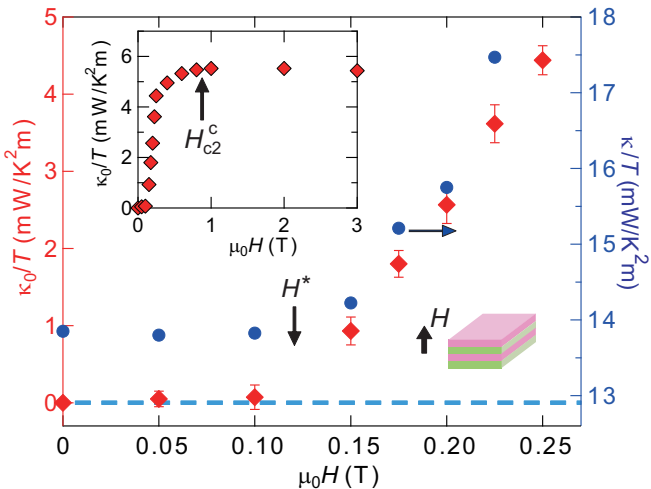}
	\caption{(Color online) Field dependence of the residual linear term $\kappa_{0}/T$ (red diamonds) and $\kappa/T$ at $T$ = 170 mK (blue circles) for {\boldmath $H$}$\parallel c$. The residual term $\kappa_{0}(H)/T$ is determined by fitting $\kappa(T)/T$ data at different fields below 0.3\,K with $\kappa_{0}/T+aT^{1.3}$, where $\kappa_{0}/T$ and $a$ are fitting parameters. Both $\kappa_{0}/T$ and $\kappa/T$ at $T$ = 170 mK are independent of magnetic field up to $H^*$. Inset shows $\kappa_{0}/T$ in an extended field range. $H_{c2}^c$ is the superconducting upper critical field along c-axis.}
\end{figure}

This conclusion is strongly supported by thermal conductivity in magnetic field applied perpendicular to the 2D plane ({\boldmath $H$}$\parallel c$).  Figure\,3 shows the $H$-dependence of $\kappa_0(H)/T$, which is obtained by the extrapolation to $T\rightarrow 0$ at each field value shown in Fig.\,2. We fitted the data in finite magnetic fields by using $\alpha=2.3$  in the same temperature range as $H=0$ ($T<$300\,mK).   As shown in Fig.\,3, $\kappa_0(H)/T$ is field independent at low field up to $\mu_0H^*\approx 0.12$\,T. In Fig.\,3, $\kappa(H)/T$ at $T$ = 170\,mK is also shown.  We stress that since $\kappa(H)/T$ at $T$ = 170 mK is also field independent up to $\sim H^*$, the field independence of $\kappa_0(H)/T$ at low-$H$ is not caused by the choice of $\alpha$.  Above $H^*$,  $\kappa_0(H)/T$ increases and becomes field independent in the normal state above $\mu_0H_{c2}^c\approx 0.8$\,T.  In the normal state,  $\kappa_0(H)/T$ is 0.0055 W/K$^2$m, which is in agreement with the value expected from the Wiedemann-Franz law.  Here we comment on the influence of Nd ions on the thermal conduction.  Since Nd ions are in the paramagnetic state in the present temperature range, magnetic excitations do not carry the heat.  Moreover the fact that $\kappa/T$ is independent of magnetic field in the low-field regime of the superconducting state  and in the normal state above $H_{c2}$ indicate that the heat conduction is not influenced by magnetic excitations.

It is well established that there is an essential difference in the field dependence of the thermal conductivity between fully gapped and nodal superconductors~\cite{Matsuda-JPhys06}.   In the former, all the quasiparticles states are bound to vortex cores and, therefore, the applied magnetic field hardly affects the thermal conduction except for the vicinity of upper critical field.  By contrast, in the latter, the heat transport is dominated by delocalized quasiparticles.  In the presence of a supercurrent with a velocity ${\bm v}_s$ around the vortices induced by magnetic field, energy of a quasiparticle with momentum {\boldmath $p$} is Doppler shifted relative to the superconducting condensate by  $E({\bm p})\rightarrow E({\bm p})-{\bm v}_s\cdot {\bm p}$.  The Doppler shift gives rise to an initial steep increase of $\kappa(H)/T\propto \sqrt{H}$ for line nodes and $\kappa(H)/T\propto H \log H$ for point nodes.  

Since $H^*$ is much larger than the lower critical field, $H^* \gg H_{c1}$, which is estimated to be $\mu_0H_{c1}=\frac{\Phi_0}{4\pi \lambda(0)^2}\ln(\frac{\lambda(0)}{\xi_{ab}})\approx 3$\,mT, the field-independent $\kappa_0/T$ is not due to the absence of flux penetration. Here, $\Phi_0$ is the flux quantum, $\lambda(0)\approx447$\,nm is the zero temperature in-plane penetration length \cite{Jiao-JPhys15} and $\xi_{ab}=\sqrt{\Phi_0/(2\pi \mu_0H_{c2}^c)}=18$\,nm is the in-plane coherence length. Therefore the observed field-insensitive thermal conductivity at low field indicates that the delocalized quesiparticles are not excited by magnetic field at least up to $H\approx H^*\approx 0.2H_{c2}^c$. These results lead us to conclude the absence of any kind of nodes in the gap function of  NdO$_{0.71}$F$_{0.29}$BiS$_{2}$.  In typical $s$-wave superconductors, such as Nb, $\kappa(H)/T$ increases steeply only in the vicinity of upper critical field~\cite{Lowell-JLTP70}.   Therefore the increase of $\kappa_0(H)/T$ above $H^*$  well below $H_{c2}$ suggests large modulation of the superconducting gap along the 2D Fermi surface. 

Finally we discuss the mechanism of the superconductivity in NdO$_{0.71}$F$_{0.29}$BiS$_{2}$.   As shown  in Fig.\,1(b), no discernible anomaly is observed in  $\kappa/T$ at $T_c$.  In the superconductors with strong electron correlation effect, including cuprates~\cite{Yu-PRL92}, iron-pnictides~\cite{Kasahara-PNAS14} and heavy fermions~\cite{izawa:prl-01,movshovich:prl-01,tanatar-prl05}, the striking enhancement of the thermal conductivity just below $T_c$ is often observed.  This enhancement is caused by the strong suppression of the quasiparticle inelastic scattering rate due to the formation of superconducting gap, which overcomes reduction of the quasiparticle density of states.  Therefore the electron correlation effect in the present compound is not strong.  Moreover, when electron-phonon coupling is strong, the  enhancement of the thermal conductivity below $T_c$ is also often observed by the enhancement of the phonon mean free path due to the gap formation.     Thus electron-phonon coupling is also not strong.   

Until now, several superconducting gap structures, including conventional $s$-, sign-reversing $s$-, spin triplet $p$-, and $d$-wave symmetries have been proposed theoretically for the BiS$_2$-based superconductors ~\cite{Usui-prb12,Wan-prb13,Yildirim-prb13,Liang-FP14,Agatsuma-JMMM16,Yang-prb13}.  Since there is no hole pocket around the $\Gamma$-point as shown in the inset of Fig.\,1(b),  we do not discuss sign-reversing $s$-wave symmetry.  Among $d$-wave symmetries, we can rule out $d$-wave with accidental nodes proposed in Ref.[\onlinecite{Usui-prb12}] and $d_{xy}$ symmetry.  The remaining possibilities of the unconventional symmetries are $p$ and $d_{x^2-y^2}$, which has no node in the present Fermi surface.  However, these unconventional superconductivity is unlikely because of the following reason. The in-plane mean free path $\ell$  is comparable to the in-plane coherence length $\xi_{ab}=\sqrt{\Phi_0/(2\pi H_{c2}^c)}=18$\,nm.  In fact, $\ell$ is estimated to be $\sim 30-50$\,nm by using the relation $\ell=(\mu_0\lambda(0)^2v_{\rm F})/\rho_0$, where Fermi velocity $v_F$ is reported to be $v_F=0.95 \times 10^6$\,m/s by angle resolved photo emission spectroscopy \cite{Ye-prb14} and $\rho_0$=500$\sim$800$\mu\Omega$cm of this study and the previous reports~\cite{Nagao-jpsj13,Liu-EPL14}.     This robustness of the superconductivity against the impurity appears to be at odds with the unconventional pairing symmetries.  These considerations lead us to conclude that  NdO$_{0.71}$F$_{0.29}$BiS$_{2}$ is likely to be a conventional $s$-wave superconductor.
 	
In summary, to clarify  the superconducting gap structure of  NdO$_{0.71}$F$_{0.29}$BiS$_{2}$,  we performed thermal conductivity  measurements down to 100\,mK.   Thermal conductivity shows no discernible anomaly at $T_c$, suggesting that both of the electron correlation and electron-phonon coupling effects are not strong.   The absence of residual  thermal conductivity indicates that a residual normal fluid, which is expected for nodal superconductors with impurities, is absent.   The magnetic field hardly affects the thermal conductivity up to $\sim 0.2 H_{c2}^c$, indicating the absence of Doppler shifted quasiparticles.  These results provide evidence that NdO$_{0.71}$F$_{0.29}$BiS$_{2}$ is a fully gapped superconductor. Notably, the estimated mean free path is comparable to the superconducting coherence length, indicating that the superconductivity is robust against impurity.  Based on these results, we conclude a conventional $s$-wave superconducting state in  NdO$_{0.71}$F$_{0.29}$BiS$_{2}$. This puts a strong constraint on the theory of the superconductivity of BiS$_2$-based layered compound, whose electronic and crystal structures bear some resemblance to Fe-based superconductors.

We thank  S. Kasahara, Y. Kasahara, Y. Mizuguchi, Y. Ota, T. Shibauchi and S. Shin for useful discussions.  This work was supported by Grants-in-Aid for Scientific Research (KAKENHI) from Japan Society for the Promotion of Science (JSPS), and by the `Topological Quantum Phenomena' (No. 25103713) Grant-in-Aid for Scientific Research on Innovative Areas from the Ministry of Education, Culture, Sports, Science and Technology (MEXT) of Japan.

\end{document}